\documentclass[pra,aps,superscriptaddress,showpacs,tightenlines,amsmath]{revtex4}
\usepackage{epsfig}
\usepackage{amsmath,amssymb}

\begin{document}

\title{Characterization process of emission sources of spin entangled pairs with several species}

\author{Francisco Delgado}
\email{fdelgado@itesm.mx}
\affiliation{Mathematics and Physics Department, Quantum Information Processing Group, Tecnologico de Monterrey, Campus Estado de Mexico, Atizapan, Estado de Mexico, CP. 52926, Mexico}

\date{\today}

\begin{abstract}
Normally, sources of entangled pairs generate several species of them. This work proposes a characterization algorithm for relatively general bipartite entangled states, generating several standard Bell states with controlled population as output. 


\end{abstract}

\maketitle

\section{Introduction}
Quantum entanglement is a fundamental resource for quantum information and quantum communication processes. Theoretical developments in these fields normally assume well delimited and stable resources as inputs, not susceptible of distortion or further complexity. That is not always the case, several recent developments in experimental generation of spin entangled pairs, as \cite{saraga1, saraga2} in quantum dots and electron gases, or as \cite{morton1, simmons1} in solid state spin ensembles, shows that normally, several species of entangled pairs are produced at the time. This resources should be characterized because they are useful for quantum computer implementations \cite{kane1}.

In addition, not only the user does not know which the produced state is, normally some kind of distortion is introduced before he has access to it. Heisenberg interactions between qubits can produce distortion on entangled pairs generated for engineering purposes (e. g.  Quantum Computation or Quantum Cryptography). The presence of parasite magnetic fields modifies the expected properties and behavior for which the pair was intended. In this sense, different kinds of quantum entangled pairs driven by a Heisenberg Hamiltonian with an additional inhomogeneous magnetic field become distorted. For these reasons, characterization of emission source should be carried out and control processes are necessary before and after of it, in order to have an useful emission as input for further quantum processes.

\section{Description of states generated by an emission source of entangled pairs}
The present problem is stated as following: a quantum system generates several kinds of entangled states with different probabilities, as a superposition of them. After of their creation, by some time $t$ before to be used, the presence of parasite magnetic fields distorts the pair driven by the Hamiltonian: $H= -J \vec{\mathbf{\sigma}}_1 \cdot \vec{\mathbf{\sigma}}_2+B_1 {\sigma_1}_z +B_2 {\sigma_2}_z$. In our discussion, we will use the generic superpostion of entangled states obtained from:

\begin{equation} \label{inistates}
\left| \psi_{1} \right> = \frac{1}{\sqrt{2}} (\left| \phi_{1} \right> \left| \phi_{2} \right>+\left| \eta_{1} \right> \left| \eta_{2} \right>)=\left| \beta_{00} \right>,  
\left| \psi_{2}(\theta) \right> = \frac{1}{\sqrt{2}} (\left| \varphi_{1} \right> \left| \varphi_{2} \right>-\left| \mu_{1} \right> \left| \mu_{2} \right>)=\sin \theta \left| \beta_{01} \right> - \cos \theta \left| \beta_{10} \right>
\end{equation}

\noindent generated by the orthogonal states (with $i=1,2$):

\begin{eqnarray}
\left| \phi_i \right> = \cos \frac{\theta}{2} \left| 0_i \right>+\sin \frac{\theta}{2} \left| 1_i \right>, 
\left| \eta_i \right> = \sin \frac{\theta}{2} \left| 0_i \right>-\cos \frac{\theta}{2} \left| 1_i \right> \\ \nonumber
\left| \varphi_i \right> = \sin \frac{\theta}{2} \left| 0_i \right>+\cos \frac{\theta}{2} \left| 1_i \right>, 
\left| \mu_i \right> = \cos \frac{\theta}{2} \left| 0_i \right>-\sin \frac{\theta}{2} \left| 1_i \right> 
\end{eqnarray}

The following state gives an almost general superposition constructed with last orthogonal entangled bipartite states:

\begin{equation} \label{generalstate} 
\left|\psi \right>=\alpha_1 \left|\psi_{1} \right> + \alpha_2 \sum_{i=1}^{2} p_i \left|\psi_{2}(\theta_i) \right>; |\alpha_1|^2+|\alpha_2|^2=p_{1}^{2}+p_{2}^{2}=1; p_1, p_2 \in \mathbb{R}, \theta_1+\theta_2=\pi/2
\end{equation}

Recently, some control processes are been developed for single qubits and entangled pairs \cite{branczyk1, delgado1} using traditional quantum gates and additional magnetic fields. These control mechanisms are precise, being limited only for the exact knowledge of the interaction strength between the pair of particles, or the inhomogeneity degree of the magnetic field which generates the distortion. Under these conditions, corrected pairs could have a remaining distortion of their expected form, but very close to it. For the state (\ref{inistates}), this control process gives:

\begin{eqnarray} \label{generalcontrolledstate} 
\left| \psi_{1} \right> &=& (1+i e^{2 i \pi n \delta}\sin{2 \pi n \delta})\left| \beta_{00} \right>-i e^{2 i \pi n \delta}\sin{2 \pi n \delta}\left| \beta_{10} \right>,  \\ \nonumber
\left| \psi_{2}(\theta) \right> &=& \sin \theta \left| \beta_{01} \right> - e^{2 i \pi n \delta} \cos{2 \pi n \delta} \cos \theta \left| \beta_{10} \right> + i e^{2 i \pi n \delta} \sin{2 \pi n \delta} \cos \theta \left|\beta_{00} \right>
\end{eqnarray}

\noindent where $\delta=j-Q(j)$, being $Q(j)$ a rational approximation of $j=J/(B_{-}^2+4J^2)^{1/2},B_{-}=B_1-B_2$ in agreement with \cite{delgado1}. Note that this possible remaining distortion can be generated by limited knowledge of $j$ or $B_{-}$ (in this case $\delta \approx -B_{-}^2/4J^2$), but otherwise can be deliberately generated to have control parameters in further processes which will use the state.

\section{Characterization algorithm, output states and populations} 
The use of non-local measurements \cite{delgado2} lets to detect the Bell states involved, by means of quantum circuit in Figure 1, where the measurements are made in the computational basis. In the following, we will use $\alpha_1=\cos \gamma$ and $\alpha_2=\sin \gamma$ (more  extended expressions for complex $\alpha_1$ and $\alpha_2$ could be considered with similar results). 

The post-measurement state and its probability (refered as population) are shown in Table 1, where $C=\sum_{i=1}^2 p_i \cos\theta_i$ and $S=\sum_{i=1}^2 p_i \sin \theta_i$. An interesting aspect is that although the post-measurement states does not depend on the control parameter, $n \delta$ (which is useful, because we want well known and standard states as output), the populations of output species depend on it. It suggest that experimental determination of $f_{00}, f_{01}, f_{10}$ and $f_{11}$ lets obtain some emission parameters, as $\gamma, C$ and $S$ (note that $C^2+S^2=1$), or alternatively to know the value of $n \delta$ when control fails to be faithful.   

\begin{table}[htb]
 \centering \caption{Summary of the application of characterization process to the input state $\left|\psi \right>$.}
\begin{tabular}{ccl}
    \hline
    Measurement outcome: $\left|i_3 j_4 \right>$ & $\left|\psi' \right>$ & Probability \\
    \hline
    $\left|0_3 0_4 \right>$ & $\left| \beta_{00} \right>$ & $f_{00} \equiv \cos^2\gamma \cos^2 2 \pi n \delta + C^2 \sin^2 \gamma \sin^2 2 \pi n \delta $ \\
    $\left|0_3 1_4 \right>$ & $\left| \beta_{01} \right>$ & $f_{01} \equiv S^2 \sin^2\gamma $ \\
    $\left|1_3 0_4 \right>$ & --- & $f_{10} \equiv 0$ \\
    $\left|1_3 1_4 \right>$ & $\left| \beta_{10} \right>$ & $f_{11} \equiv C^2 \sin^2\gamma \cos^2 2 \pi n \delta + \cos^2 \gamma \sin^2 2 \pi n \delta $ \\
    \hline
   \end{tabular}
    \end{table}

Otherwise, if $n \delta$ is deliberately constructed, it lets to control the relative population for two of the three resulting species by noting that if $S$ fulfills:

\begin{equation} \label{sennd} 
S^2=\frac{1-f_{00}-f_{11}}{\sin^2\gamma} \Rightarrow \sin^2 2 \pi n \delta = \frac{\cos^2 \gamma -f_{00}}{\cos 2 \gamma +1-f_{00}-f_{11}}
\end{equation}

\noindent there are ample posibilities to control the populations despite last restrictions for $f_{00}, f_{11}$ and $\gamma$. Figure 2 shows the region for these parameters in which the control it is possible. By example, taking the $\gamma$ parameter which characterize the emission source, there are a two-dimensional slice region to select $f_{00}$ and $f_{11}$ inside which last restrictions are fulfilled to give a physical $n \delta$ to obtain the desired populations.

\section{Conclusions} 
Accute control of quantum entanglement sources are actually required by controlling their output resources preventing them from decoherence and giving adequate properties under design to use them for quantum engineering purposes. The actual characterization process of spin entangled pairs sources combined with previous control to prevent distortion generated by magnetic fields lets manipulate the relative rates of specific Bell states species originally generated for that source. In addition, the knowledge of some parameters, which characterize it, can be obtained with the quantum algorithm presented here by measuring natural relative populations. This process requires only to know basic properties of the source as particles' nature and average distances involved. Nevertheless, the process presented is limitated to control only two species of the three species available, so improvements should let manipulate all species involved and include other ones.

\section*{FIGURE CAPTIONS}
\subsection*{Figure 1}
Quantum circuit for characterization process.

\subsection*{Figure 2}
Control region in which it is possible select the population of species $f_{00}$ and $f_{11}$ as function of $\gamma$ parameter. The plane shows the case for $\gamma=\pi/4$; inside of two dimensional cut slice it is possible to select desired values for those populations with the correct selection of $n \delta$ parameter in agreement with (\ref{sennd}). This control process have more freedom when $\gamma=\pi/2$, but $f_{01}=0$.

\end{document}